\def\blankline{\vskip \baselineskip}

\def\degrees{{$^\circ$}}
\def\half{{\leavevmode\kern.1em\raise.5ex\hbox{\the\scriptfont0 1}\kern-.1em
/\kern-.1em\lower.25ex\hbox{\the\scriptfont0 2}}} 
\def\quarter{{\leavevmode\kern.1em\raise.5ex\hbox{\the\scriptfont0 1}\kern-.1em
/\kern-.1em\lower.25ex\hbox{\the\scriptfont0 4}}}

\def\spose#1{\hbox to 0pt{#1\hss}} 
\def\ltsim{\mathrel{\spose{\lower.5ex\hbox{$\mathchar"218$}}
     \raise.4ex\hbox{$\mathchar"13C$}}}
\def\gtsim{\mathrel{\spose{\lower.5ex \hbox{$\mathchar"218$}}
     \raise.4ex\hbox{$\mathchar"13E$}}}
\def\gtlt{\mathrel{\spose{\lower.5ex\hbox{$\mathchar"13E$}}
     \raise.5ex\hbox{$\mathchar"13C$}}}

\def\pmb#1{\setbox0=\hbox{$#1$}%
  \kern-0.25em\copy0\kern-\wd0
  \kern.05em\copy0\kern-\wd0
  \kern-0.025em\raise.0433em\box0}
\def\spmb#1{\setbox1=\hbox{${\scriptstyle #1}$}%
  \kern-0.25em\copy1\kern-\wd1
  \kern.05em\copy1\kern-\wd1
  \kern-0.025em\raise.0433em\box1}

\def\today{\count99=\day
           \ifnum\count99>20 \count98=\day
                             \divide\count98 by 10
                             \multiply\count98 by 10
                             \advance\count99 by -\count98 \fi
           \number\day\ifcase\count99 th\or st\or nd\or rd\else th\fi
           ~\ifcase\month none\or January\or February\or March\or April\or
                  May\or June\or July\or August\or September\or October\or
                  November\or December\fi
           ~\number\year}

\long\def\Ignore#1{\relax}%

\newdimen\digitwidth
\setbox0=\hbox{0}
\digitwidth=\wd0
%
%

\def\eg{{\it e.g.}}
\def\etal{{\it et al.}}

\def\ie{{\it i.e.}}

\def\viz{{\it viz.}}

\newcount\linespacingstep
\newcount\linespacing
\linespacingstep=1
\def\multiplelines{\linespacing=\linespacingstep
   \advance\linespacing by 2
   \multiply\normalbaselineskip by \linespacing
   \advance\normalbaselineskip by 2pt 
   \divide\normalbaselineskip by 3}

\font\fiverm=cmr5	\font\sixrm=cmr6	\font\sevenrm=cmr7
\font\eightrm=cmr8 	\font\ninerm=cmr9	\font\tenrm=cmr10
\font\elevenrm=cmr11 	\font\twelverm=cmr12 	\font\seventeenrm=cmr17

\font\fivei=cmmi5 	\font\sixi=cmmi6 	\font\seveni=cmmi7
\font\eighti=cmmi8 	\font\ninei=cmmi9 	\font\teni=cmmi10
\font\eleveni=cmmi11    \font\twelvei=cmmi12

\font\fivesy=cmsy5  	\font\sixsy=cmsy6 	\font\sevensy=cmsy7
\font\eightsy=cmsy8 	\font\ninesy=cmsy9 	\font\tensy=cmsy10
\font\elevensy=cmsy10 at 11pt \font\magnifiedtensy=cmsy10 at 12pt

\font\fivebf=cmbx5  	\font\sixbf=cmbx6 	\font\sevenbf=cmbx7
\font\eightbf=cmbx8 	\font\ninebf=cmbx9 	\font\tenbf=cmbx10
\font\elevenbf=cmbx11   \font\twelvebf=cmbx12

	\font\eightit=cmti8 	\font\nineit=cmti9
\font\tenit=cmti10	\font\elevenit=cmti11	\font\twelveit=cmti12

\font\eightsl=cmsl8 	\font\ninesl=cmsl9	\font\tensl=cmsl10
\font\elevensl=cmsl11   \font\twelvesl=cmsl12

\font\eighttt=cmtt8 	\font\ninett=cmtt9 	\font\tentt=cmtt10
\font\eleventt=cmtt11   \font\twelvett=cmtt12

\font\tenex=cmex10

\def\eightpoint{\def\rm{\fam0\eightrm}
\textfont0=\eightrm\scriptfont0=\sixrm\scriptscriptfont0=\fiverm
\textfont1=\eighti\scriptfont1=\sixi\scriptscriptfont1=\fivei
\textfont2=\eightsy\scriptfont2=\sixsy\scriptscriptfont2=\fivesy
\textfont3=\tenex\scriptfont3=\tenex\scriptscriptfont3=\tenex
\textfont\itfam=\eightit\def\it{\fam\itfam\eightit}
\textfont\slfam=\eightsl\def\sl{\fam\slfam\eightsl}
\textfont\ttfam=\eighttt\def\tt{\fam\ttfam\eighttt}
\textfont\bffam=\eightbf\scriptfont\bffam=\sixbf
\scriptscriptfont\bffam=\fivebf\def\bf{\fam\bffam\eightbf}
\normalbaselineskip=9pt
\ifnum\linespacingstep>1\multiplelines\fi
\setbox\strutbox=\hbox{\vrule height7pt depth2pt width0pt}
\let\sc=\sixrm\let\big=\eightbig\normalbaselines\rm}

\def\ninepoint{\def\rm{\fam0\ninerm}
\textfont0=\ninerm\scriptfont0=\sixrm\scriptscriptfont0=\fiverm
\textfont1=\ninei\scriptfont1=\sixi\scriptscriptfont1=\fivei
\textfont2=\ninesy\scriptfont2=\sixsy\scriptscriptfont2=\fivesy
\textfont3=\tenex\scriptfont3=\tenex\scriptscriptfont3=\tenex
\textfont\itfam=\nineit\def\it{\fam\itfam\nineit}
\textfont\slfam=\ninesl\def\sl{\fam\slfam\ninesl}
\textfont\ttfam=\ninett\def\tt{\fam\ttfam\ninett}
\textfont\bffam=\ninebf\scriptfont\bffam=\sixbf
\scriptscriptfont\bffam=\fivebf\def\bf{\fam\bffam\ninebf}
\normalbaselineskip=11pt
\ifnum\linespacingstep>1\multiplelines\fi
\setbox\strutbox=\hbox{\vrule height8pt depth3pt width0pt}
\let\sc=\sevenrm\let\big=\ninebig\normalbaselines\rm}

\def\tenpoint{\def\rm{\fam0\tenrm}
\textfont0=\tenrm\scriptfont0=\sevenrm\scriptscriptfont0=\fiverm%
\textfont1=\teni\scriptfont1=\seveni\scriptscriptfont1=\fivei%
\textfont2=\tensy\scriptfont2=\sevensy\scriptscriptfont2=\fivesy%
\textfont3=\tenex\scriptfont3=\tenex\scriptscriptfont3=\tenex%
\textfont\itfam=\tenit\def\it{\fam\itfam\tenit}%
\textfont\slfam=\tensl\def\sl{\fam\slfam\tensl}%
\textfont\ttfam=\tentt\def\tt{\fam\ttfam\tentt}%
\textfont\bffam=\tenbf\scriptfont\bffam=\sevenbf%
\scriptscriptfont\bffam=\fivebf\def\bf{\fam\bffam\tenbf}%
\normalbaselineskip=12pt%
\ifnum\linespacingstep>1\multiplelines\fi
\setbox\strutbox=\hbox{\vrule height8.5pt depth3.5pt width0pt}%
\let\sc=\eightrm\let\big=\tenbig\normalbaselines\rm}

\def\elevenpoint{\def\rm{\fam0\elevenrm}
\textfont0=\elevenrm\scriptfont0=\eightrm\scriptscriptfont0=\sixrm%
\textfont1=\eleveni\scriptfont1=\eighti\scriptscriptfont1=\sixi%
\textfont2=\elevensy\scriptfont2=\eightsy\scriptscriptfont2=\sixsy%
\textfont3=\tenex\scriptfont3=\tenex\scriptscriptfont3=\tenex%
\textfont\itfam=\elevenit\def\it{\fam\itfam\elevenit}%
\textfont\slfam=\elevensl\def\sl{\fam\slfam\elevensl}%
\textfont\ttfam=\eleventt\def\tt{\fam\ttfam\eleventt}%
\textfont\bffam=\elevenbf\scriptfont\bffam=\eightbf%
\scriptscriptfont\bffam=\sixbf\def\bf{\fam\bffam\elevenbf}%
\normalbaselineskip=13pt%
\ifnum\linespacingstep>1\multiplelines\fi
\setbox\strutbox=\hbox{\vrule height9.5pt depth4pt width0pt}%
\let\sc=\ninerm\let\big=\elevenbig\normalbaselines\rm}

\def\twelvepoint{\def\rm{\fam0\twelverm}
\textfont0=\twelverm\scriptfont0=\eightrm\scriptscriptfont0=\sixrm
\textfont1=\twelvei\scriptfont1=\eighti\scriptscriptfont1=\sixi
\textfont2=\magnifiedtensy\scriptfont2=\eightsy\scriptscriptfont2=\sixsy
\textfont3=\tenex\scriptfont3=\tenex\scriptscriptfont3=\tenex
\textfont\itfam=\twelveit\def\it{\fam\itfam\twelveit}
\textfont\slfam=\twelvesl\def\sl{\fam\slfam\twelvesl}
\textfont\ttfam=\twelvett\def\tt{\fam\ttfam\twelvett}
\textfont\bffam=\twelvebf\scriptfont\bffam=\eightbf
\scriptscriptfont\bffam=\sixbf\def\bf{\fam\bffam\twelvebf}
\tt 
\normalbaselineskip=14pt
\ifnum\linespacingstep>1\multiplelines\fi
\setbox\strutbox=\hbox{\vrule height10pt depth5pt width0pt}
\let\sc=\eightrm\let\big=\twelvebig\normalbaselines\rm}

\twelvepoint

\vsize=22.6 true cm \hsize=17 true cm 
\clubpenalty=5000
\widowpenalty=5000

\input psfig
\hsize=15.5cm\vsize=23cm
\parskip=2pt
\pageno=155
\headline={\ifnum\pageno=155 \else
    \ifodd\pageno {\twelveit \runhead} \hfil {\twelverm \folio}
       \else {\twelverm \folio} \hfil {\twelveit \names}
    \fi
  \fi}
\nopagenumbers
\def\names{Sellwood}
\def\runhead{Spiral instabilities in N-body simulations}

\def\figone{\vbox to 17.5cm{\hsize = 12.5cm
{\psfig{file=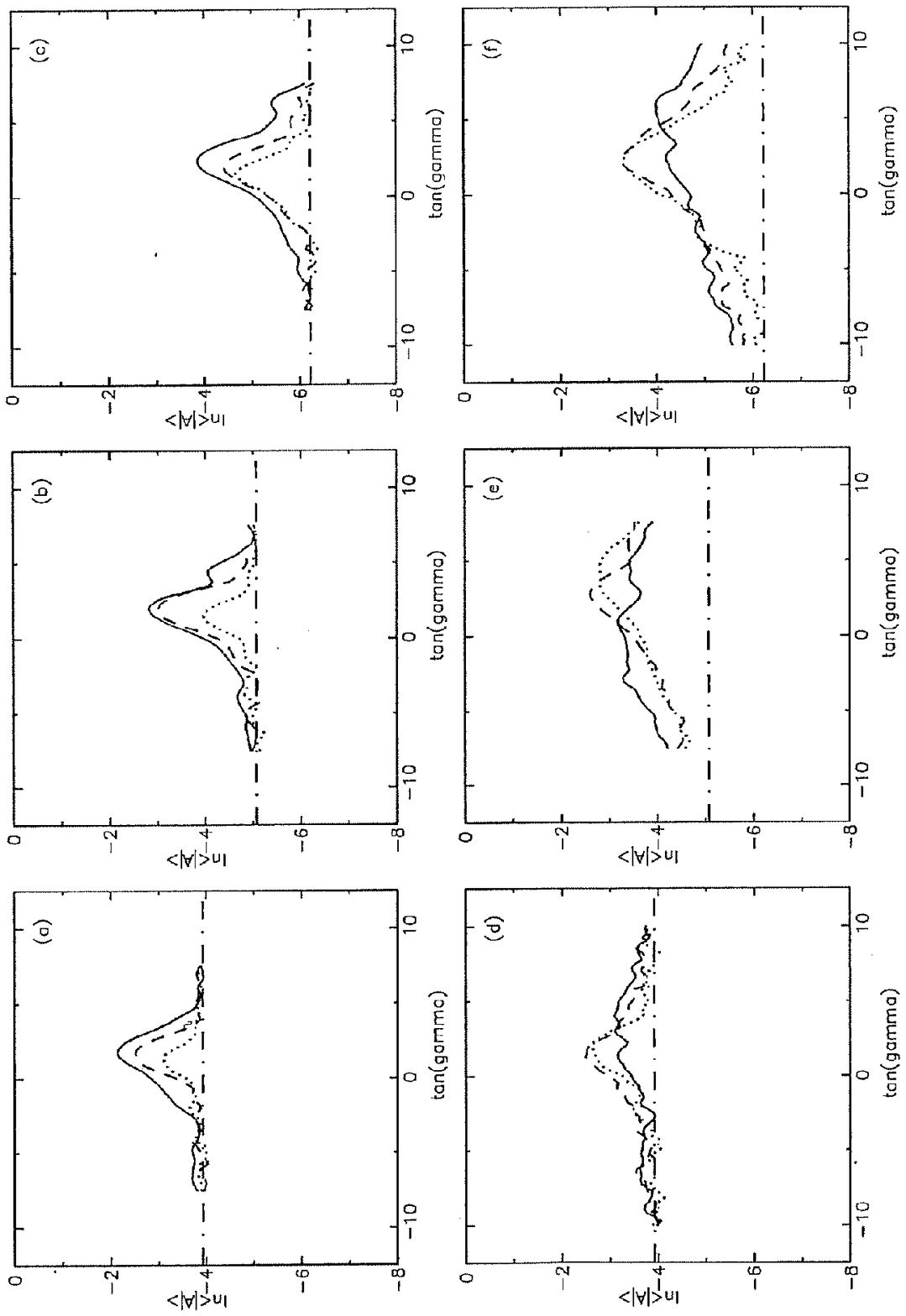,width=\hsize}
{\noindent\tenpoint{\bf Figure 1}  Time averaged values of the transform (equation 1) from two series
of experiments in which the particle number is changed.  The results for $m=2$
-- full drawn, $m=3$ -- dashed and $m=4$ -- dotted curves are shown in each
case.  The horizontal dot-dash line indicates the expectation value for $N$
randomly distributed particles.  The top row (a-c), show results from Zang
models, the bottom row from Sc models, and $N$ rises from 2K (left hand panels)
through 20K (centre panels) to 200K (right hand panels).\par}}\vfill}}

\def\figtwo{\vbox to 17.5cm{\hsize = 12.5cm{\vfill\psfig{file=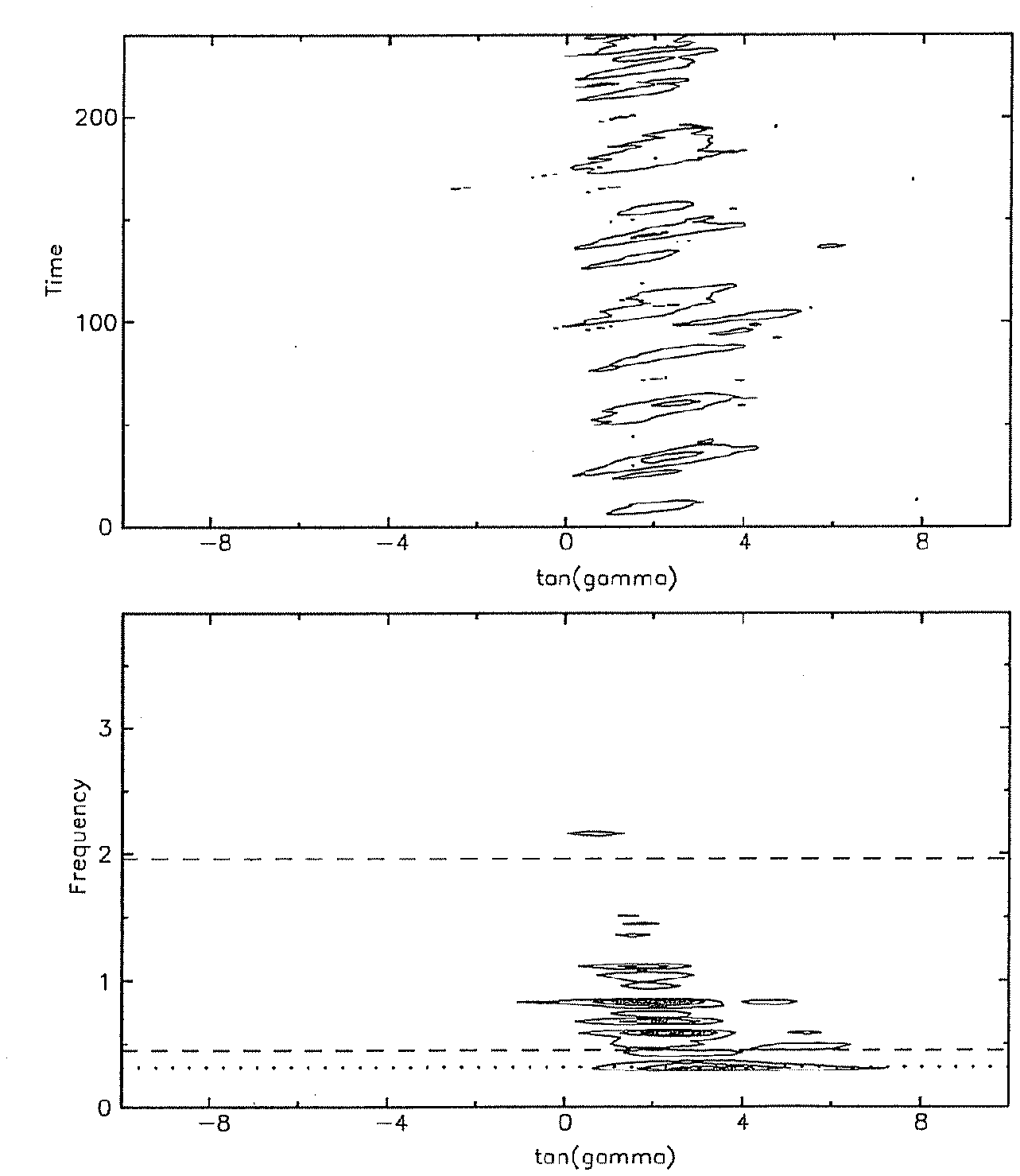,width=\hsize}
{\noindent\tenpoint{\bf Figure 2} Results from a quasi-stationary Sc model. The rotation period
at the half-mass radius in this model is 16 time units.  (a) Contours of the
$m=3$ component of $|A|$ as a function of time and inclination angle.  (b) A
power spectrum of the same data, revealing a number of coherent waves at
constant frequencies over wide ranges in pitch angles. \par}}
\noindent\hbox to 5cm{\kern-1cm\hrulefill}}}

\def\figthree{\vbox to \vsize{\vfill\psfig{file=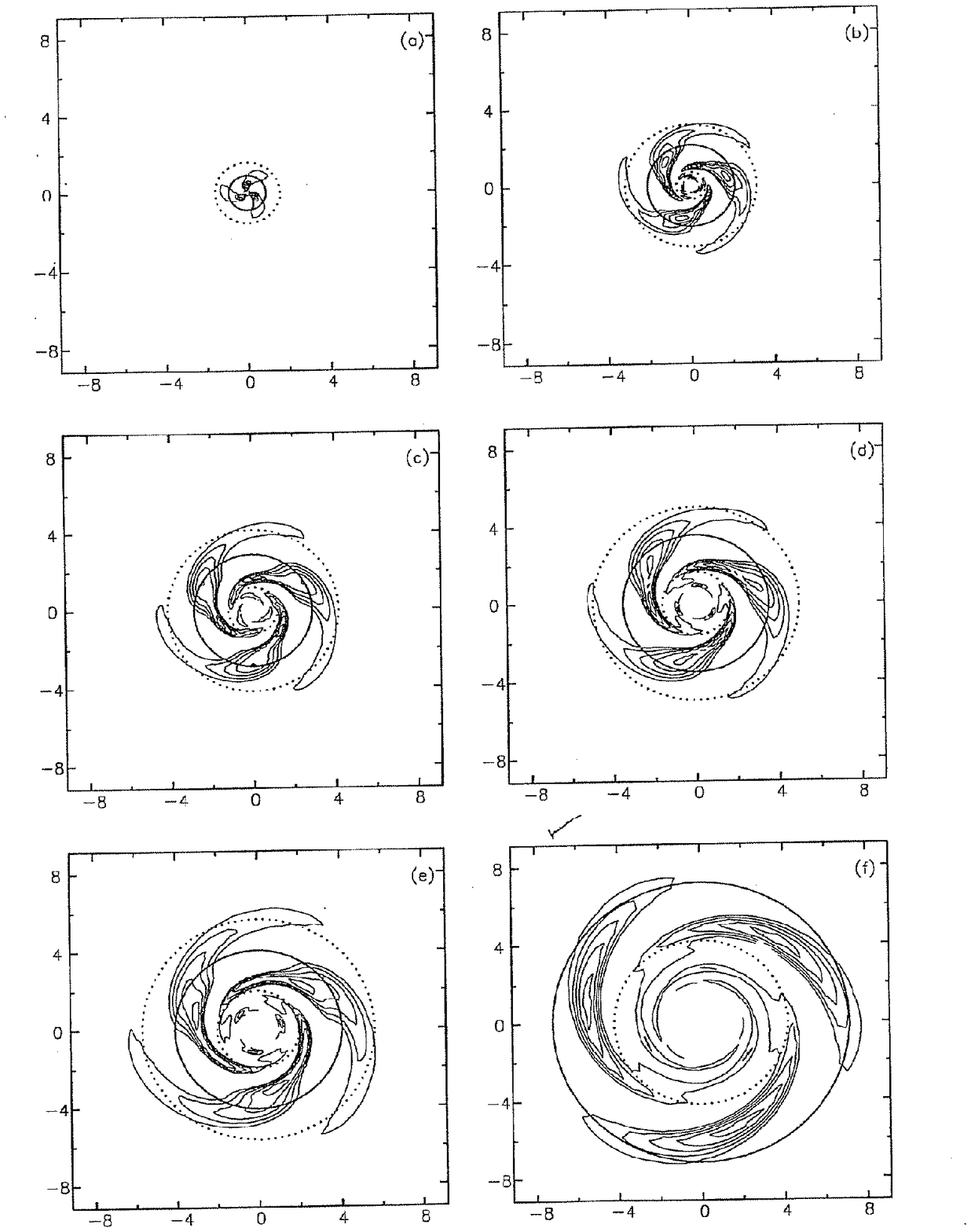,width=\hsize}
{\noindent\tenpoint{\bf Figure 3} Six representative waves from the power spectrum in Figure~3(b).  The
co-rotation circle (full drawn) and the Lindblad resonances (dotted) are
marked.  The scales are in units of the scale radius of the Sc model (SC).
\par}}\vfill}

\def\figfour{\vbox to 10cm{\hsize = 13.5cm{\psfig{file=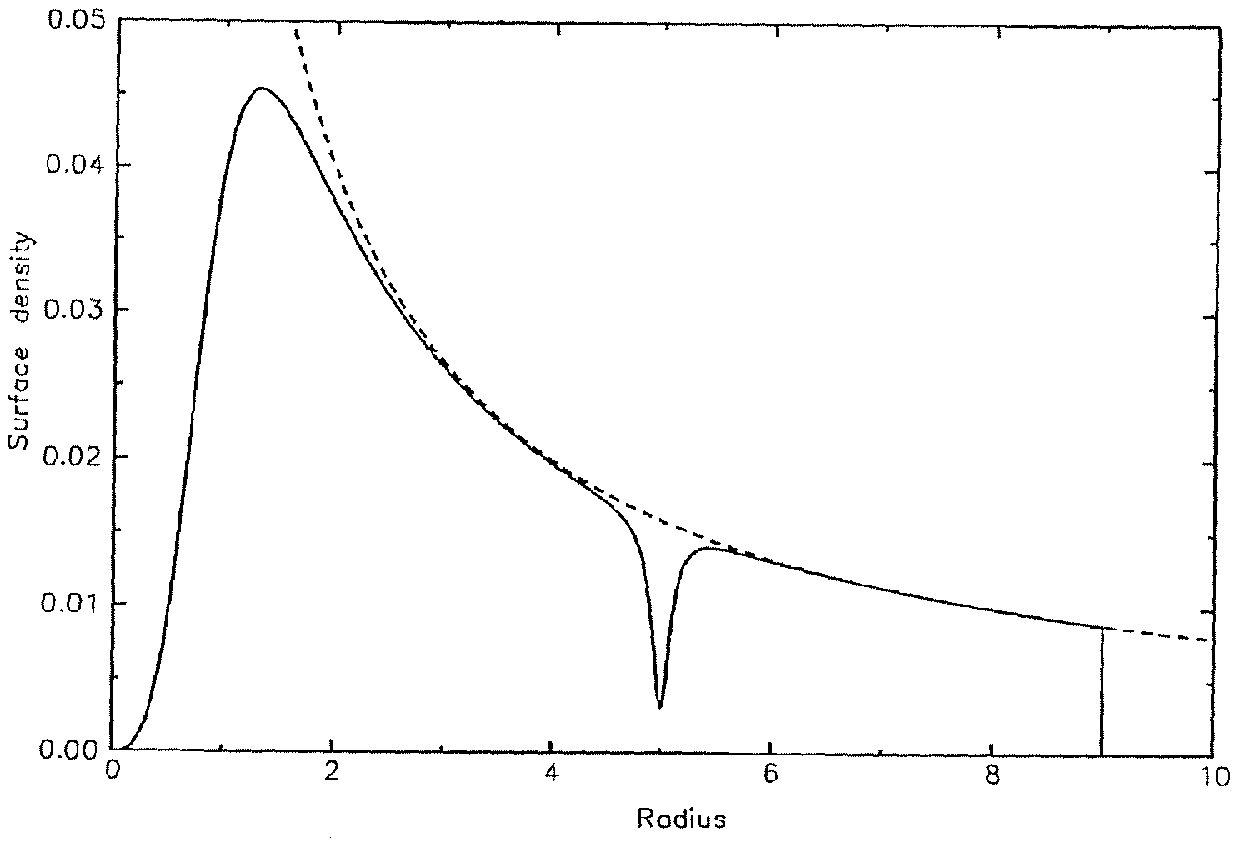,width=\hsize}\vfill
{\noindent\tenpoint{\bf Figure 4} The axisymmetric surface density distribution in the model
illustrated in Figure~5 (full drawn curve).  The dashed curve shows the basic
Mestel disc without an inner taper, groove or outer edge.  Units are in $r_0$
and $V_0^2/(Gr_0)$. \par}}
\noindent\hbox to 5cm{\kern-1cm\hrulefill}}}

\def\figfive{\vbox to \vsize{\psfig{file=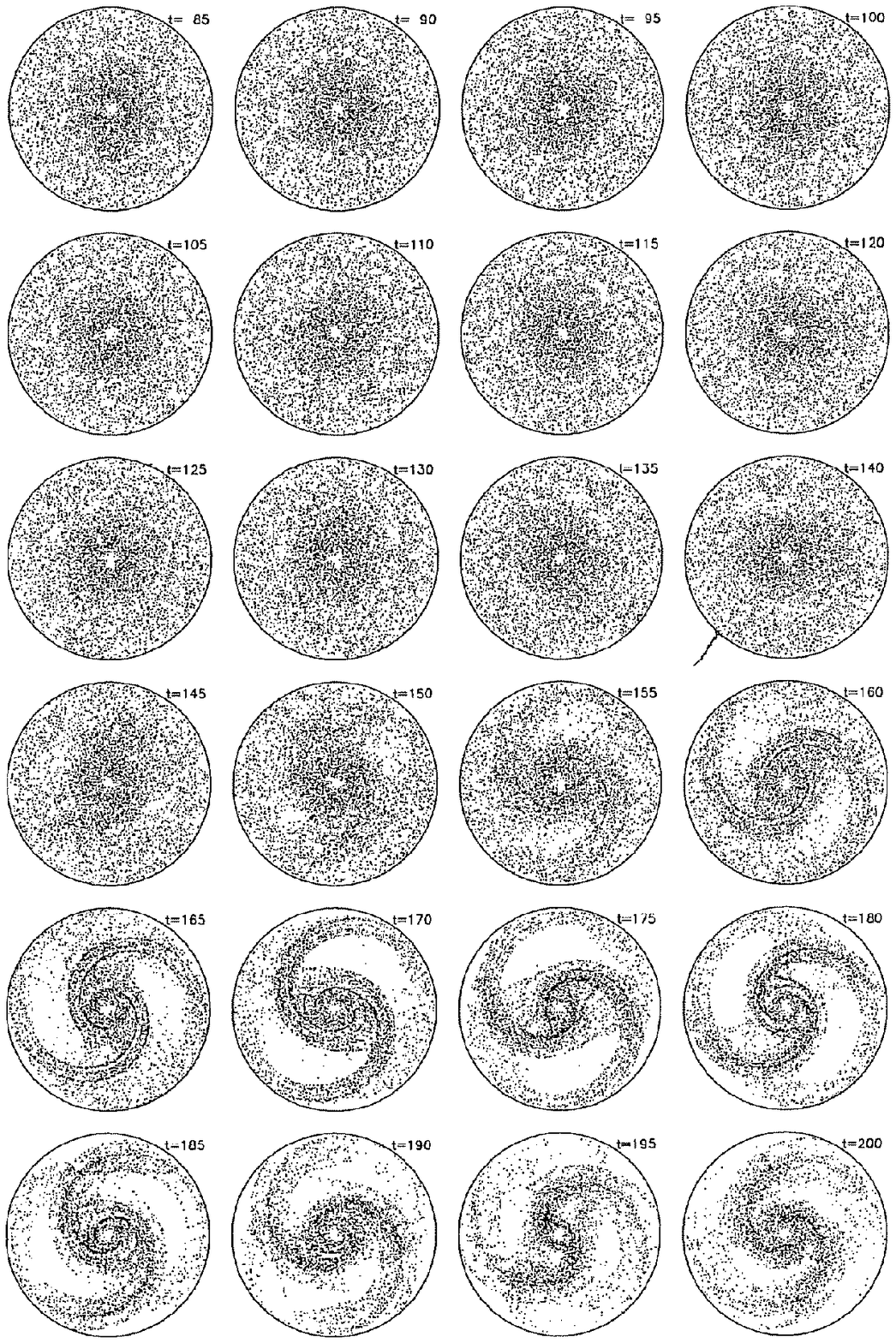,width=.95\hsize}\vfill
{\noindent\tenpoint{\bf Figure 5} Evolution of a cold disc model with a deep, narrow groove.  Times
are in units of $r_0/V_0$ -- a rotation period at the groove is therefore
$10\pi$.  Note that the early evolution of the model, during which no visible
changes occur, is omitted.  The calculation was restricted so that only $m=2$
components of the disturbance potential affected the motion of the particles.
\par}}}

\def\figsix{\vbox to 13cm{\hsize = 13.5cm{\hfil\psfig{file=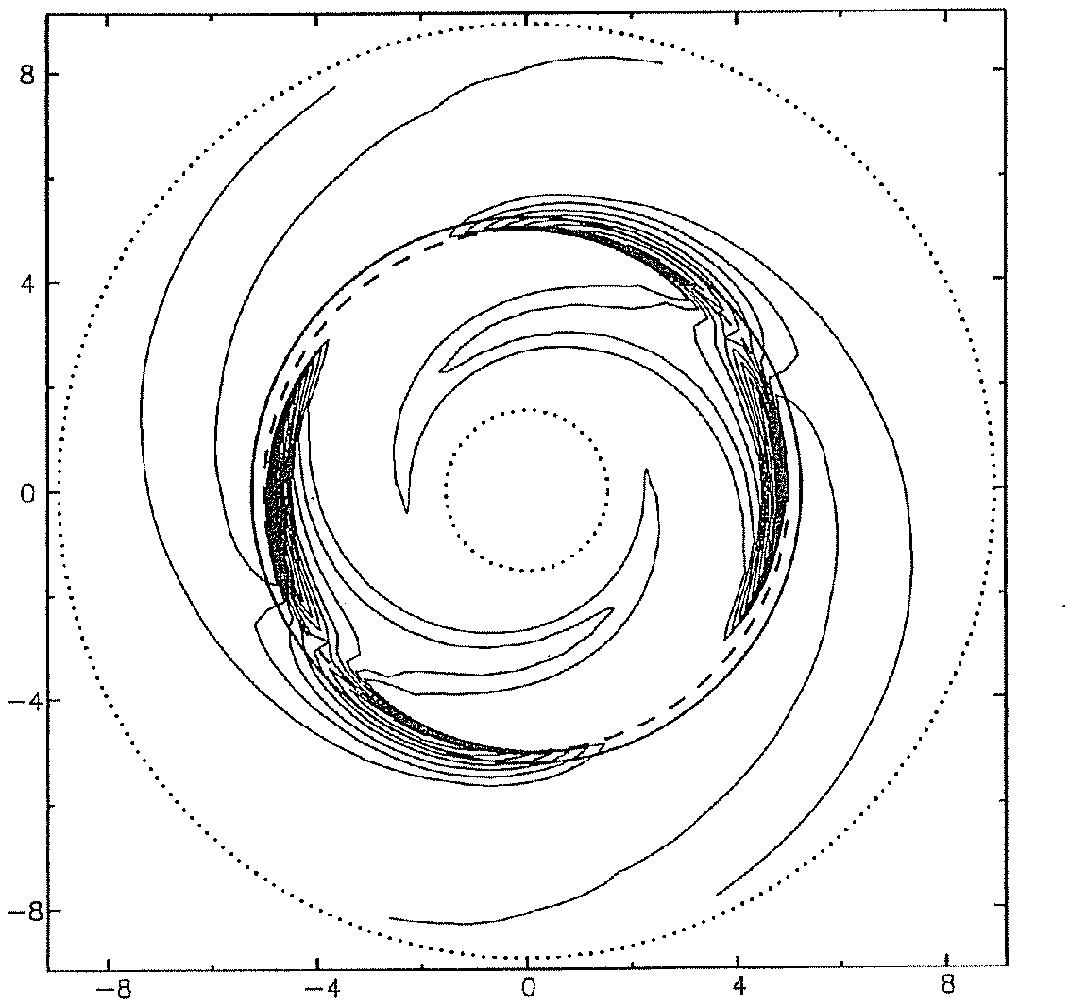,width=.9\hsize}\hfil\vfill
{\noindent\tenpoint{\bf Figure 6} The best fit exponentially growing mode to the small amplitude
evolution of the model shown in Figure~5.  Only the positive part is shown. 
The full drawn circle marks co-rotation and the dotted circles the Lindblad
resonances for the best-fit frequency.  The groove centre is marked by the
dashed circle. \par}}
\noindent\hbox to 5cm{\kern-1cm\hrulefill}}}

\def\figseven{\vbox to \vsize{\psfig{file=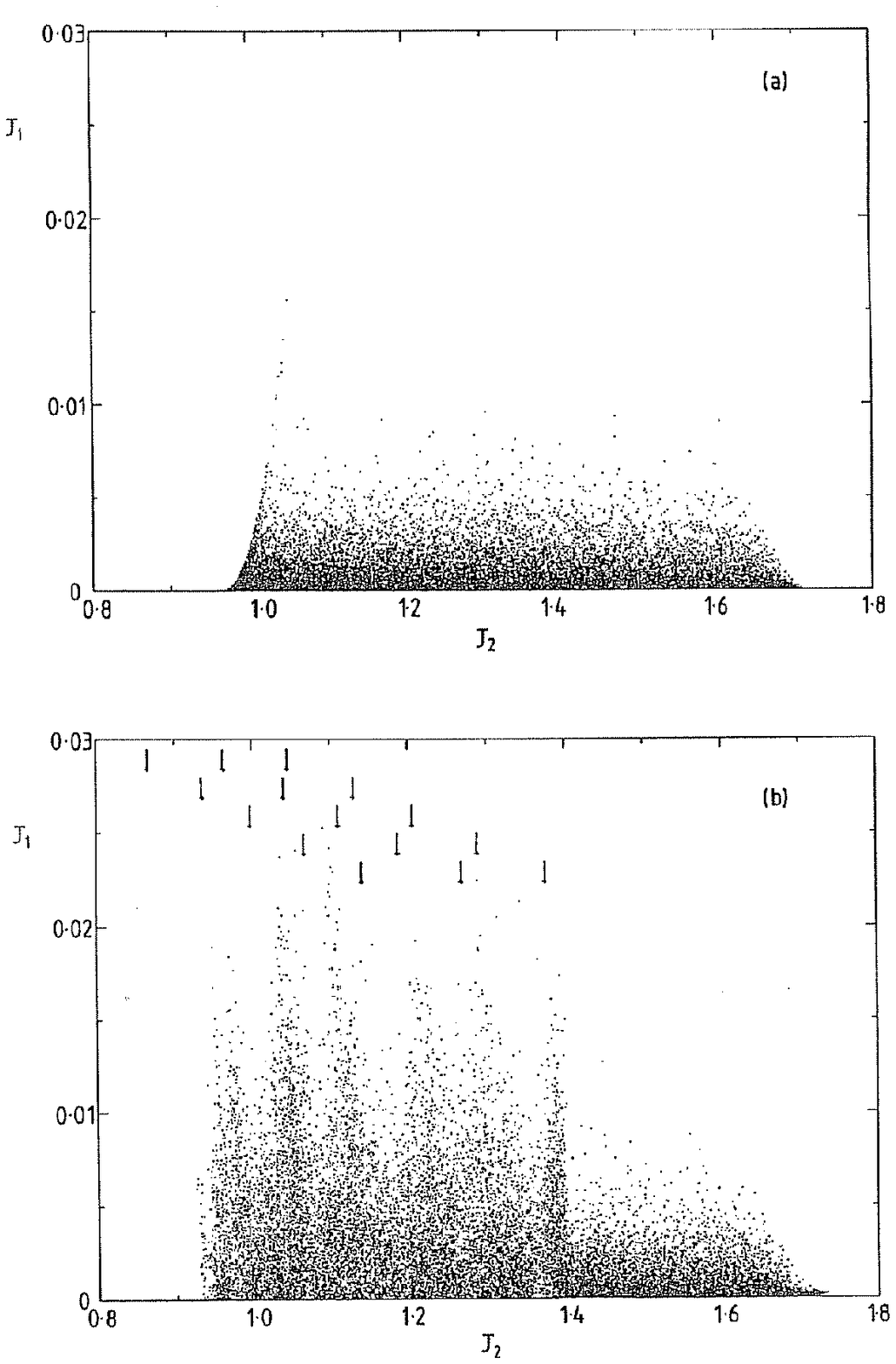,width=\hsize}
{\noindent\tenpoint{\bf Figure 7} The phase space distribution of particles in the experiment described
in Sellwood \& Lin (1989)  (a) Shows the situation at the start and (b) after
267 dynamical times.  The spread in $J_2$ (angular momentum) reflects the
radial extent of the annulus and those particles with large $J_1$ (radial
action) are on more eccentric orbits.  Only one fifth of the 100K particles used
in the simulation are plotted in each panel.  The arrows in (b) point to the
locations of co-rotation and the two Lindblad resonances for the five waves
observed in the model by time 267.\par}}\vfill}

\def\title#1{\centerline{\seventeenrm #1}}
\def\author#1{\centerline{\bf #1}}
\def\affil#1{\centerline{\tenit #1}}
\def\abstract#1{{\tenpoint \noindent{\bf Abstract} \hskip 1ex #1 \par}}
\def\lone#1{\goodbreak\bigskip \noindent{\bf #1} \par\medskip\noindent \ignorespaces}
\def\ltwo#1{\goodbreak\bigskip \noindent\uppercase{#1} \par\smallskip\noindent \ignorespaces}
\topinsert{\bigskip}\endinsert
\title{Spiral instabilities in N-body simulations}
\bigskip
\author{J. A. Sellwood}
\affil{University of Manchester}
\bigskip

{\narrower\abstract{$N$-body simulations of disc galaxies that display recurrent
transient spiral patterns are comparatively easy to construct, but are harder
to understand.  In this paper, I summarise the evidence from such experiments
that the spiral patterns result from a recurrent spiral instability cycle.  
Each wave starts as rapidly growing, small-amplitude instability caused by a
deficiency of particles at a particular angular momentum.  The resulting
large-amplitude wave creates, through resonant scattering, the conditions
needed to precipitate a new instability.}}

\lone{1 Plan}
The problem of spiral structure in galaxies has been worked on for many years
but progress has been painfully slow.  Most effort has been directed towards
the development of an analytical (or at least semi-analytical) approach and
many aspects of the problem have been discovered (see Sellwood 1989 for a
review).  Here, I collect the evidence from $N$-body simulations which
indicates that the structure is continuously variable and results from a
recurrent cycle of spiral instabilities. 

A subsidiary purpose of this paper, is to convince the reader of the advantages
of using $N$-body simulations {\it in tandem\/} with approximate analytic
treatments.  Without a close comparison of this nature, each separate approach
is much less powerful; the limitations of the $N$-body experiments remain
unquantified and the validity of the approximations in the analytic approach
cannot be assessed. 

The paper is divided into three distinct sections.  In \S2, I discuss
swing-amplified noise in global simulations, and show that the behaviour in the
Mestel ($V=$~const.) disc is very similar to that reported by Toomre (\eg\ this
conference) for simulations in the shearing sheet.  However, other more
realistic models display much larger amplitude structure and wave coherence
than can be accounted for by swing-amplified noise alone. 

At first sight, the instability caused by a groove in a disc, which is
described in \S3, is totally unrelated to the previous section.  The vigorous
instability provoked by such an apparently arbitrary feature leads to a large
amplitude spiral pattern.  I present a simplified local analytic treatment
which yields a rough prediction for the eigenfrequency. 

The third part of this story describes an experiment which suggests how the
two previous sections might be related.  The coherent waves discussed briefly
in \S4 appear to result from a {\it recurrent cycle\/} of groove instabilities.

\lone{2 More than particle noise}
\vskip -2.5ex
\ltwo{2.1 Zang discs}
The stability properties of the infinite Mestel (1963) disc are quite
remarkable.  The disc has the surface density $$ 
\Sigma(r) = {qV_0^2 \over 2\pi Gr}, 
$$ which gives an exactly flat rotation curve all the way from the axis of
rotation.  Here, $V_0$ is the orbital speed for circular motion at any radius
$r$ which arises from this mass distribution when $q=1$.  Though the surface
density is singular at the centre, the mass within any radius is, of course,
finite. 

Zang (1976) examined the global stability properties of centrally cut-out, but
otherwise full-mass, models having this density distribution and found, rather
surprisingly, that the most persistently unstable modes occurred for $m=1$. 
The $m=2$ modes could be completely suppressed if the velocity dispersion were
comparatively modest and the central cut-out not too abrupt (see also Toomre
1977).  The central cut out he used took the form of a taper in angular
momentum, $J$, of the distribution function: $T(J) = [1 + (r_0V_0/J)^n]^{-1}$,
where $r_0$ is some scale radius and the index $n \ltsim 2$ for stability.  It
now seems clear that the instabilities Zang found in models with $n>2$ were
edge-related modes.

In largely unpublished subsequent work, referred to in Toomre (1981), Zang and
Toomre extended the study to include discs for which $q<1$, \ie\ embedded
within a halo that did not alter the shape of the rotation curve.  They found
that a half-mass disc model ($q=0.5$) could be {\it completely stable\/}
to all global modes, provided the central taper was not too abrupt ($n \leq
4$).  The radial velocity dispersion in their stable model $\sigma_u
=0.2835V_0$ which gives $Q=1.5$ for all $r \gg r_0$. 

The absence of instabilities makes this model an ideal test-bed in which to
study, in a global context, a variety of phenomena that have been studied in
the (local) shearing sheet model.  Indeed Toomre (1981) already gives a
memorable illustration of swing-amplification that is far more eye-catching
than any diagram produced in an analysis of the shearing sheet, even though it
was first discovered in that context some fifteen years earlier (Goldreich \&
Lynden-Bell 1965, Julian \& Toomre 1966, hereafter JT). 

\medskip
\ltwo{2.2 Swing-amplified noise}
I have therefore run a series of simulations of this model which illustrate
swing-amplified particle noise in a global context.  These experiments are the
logical counterparts to Toomre's simulations in the local shearing sheet model
(a preliminary report of which is given in this volume).  These are essentially
empirical measurements of polarisation in a disc of particles characterised by
$Q=1.5$.  As quiet starts would obviously be inappropriate for such an
investigation, the initial azimuthal coordinates were chosen randomly. 

Unfortunately, it is necessary, for obvious computational reasons, to truncate
the disc at some outer radius also.  I have not yet succeeded in doing this
without provoking at least a mild $m=2$ instability at the outer edge.  Though
the simulations are not completely stable, therefore, the growth of the outer
edge mode is slow enough that the behaviour for the first hundred dynamical
times is almost completely consistent with swing-amplified particle noise, as
can be seen from the following. 

In order to measure the amplitude of density variations, I formed the summation
$$ 
A(m, \gamma, t) = {1 \over N} \sum_{j=1}^N \exp[im(\theta_j + \tan\gamma\ln
r_j)],  \eqno(1)
$$
at intervals during the simulations.  Here, $N$ is the number of particles
and $(r_j, \theta_j)$ are the coordinates of the $j$th particle at time $t$. 
The resulting complex coefficients $A$ are the logarithmic spiral
transformation of the particle distribution and $\gamma$ is the inclination
angle of the spiral component to the radial direction (positive for trailing
waves). 

\pageinsert{\vfill \hbox to \hsize{\hfill\figone\hfill} \vfill}\endinsert

The top line of panels in Figure~1(a--c) shows the time-averaged value of $|A|$
as a function of $\tan\gamma$ for the $m=2$, 3 \& 4 components from three
different experiments which span two decades in $N$.  The time interval, in
units of $r_0/V_0$, chosen for the average was $t=25$ to $t=100$ in each case
-- a period after the polarisation is reasonably well developed over most of
the disc and before the outer edge instability had reached any significant
amplitude. 

The leading/trailing bias is obvious in all these plots, and is largest for
$m=2$.  The horizontal dot-dash lines indicate the expectation value of $|A| (=
\sqrt{\pi/4N}$) for randomly distributed particles (Sellwood \& Carlberg 1984).
The measured amplitude on the far leading side is completely consistent with
randomly distributed particles in all three experiments.  The peaks on the
trailing side are higher than the leading signal by an approximately equal
factor in the $N=20$K and $N=200$K experiments, but noticeably less when
$N=2$K.  Spiral amplitudes in the $N=2$K experiment are so large as to be
clearly visible which probably limits the measured bias for two reasons: peak
amplitudes may well be limited by non-linear effects, and the velocity
dispersion rises very quickly -- making the disc less responsive.  [This
simulation is illustrated in Sellwood (1986).] 

Following Toomre, I have estimated the bias that should be observed using the
JT apparatus and found excellent agreement for the $m=3$ \& 4 components -- the
$m=2$ bias in these global experiments is slightly larger than predicted by
this local theory (Sellwood, in preparation). 

In conclusion, results from these three experiments of this very nearly stable
disc are quite consistent with theoretical expectations of swing-amplified
particle noise. 

\ltwo{2.3 More realistic models}
Measurements over a comparable time interval from a parallel series of
experiments with the ``Sc'' model of a disc galaxy used by Sellwood \& Carlberg
(1984, hereafter SC) are shown in the bottom row of panels in Figure~1(d--e). 
These experiments are not quite as comparable as would be desirable -- it was
necessary to use a lower mass fraction in the disc (30\%) in order to prevent a
rapid bar instability at the centre, and the initial velocity dispersion was
set so as $Q=1$ only.  These differences shift the peak response to $m=3$ and
make the discs more responsive, respectively. 

Nevertheless, the behaviour differs from that in the Zang models by more than
can be accounted for by these reasons alone.  The most obvious two differences
are that the amplitudes on the leading side in the larger $N$ experiments are
no longer consistent with randomly distributed particles and nowhere do the
signals decrease as $1/\sqrt N$.  Though the amplitudes in the 200K particle
experiment (f) are lower than those in (e), they rise throughout this time
interval (1 -- 4 rotation periods at the half-mass radius), attaining values
no different from those in the lower $N$ experiments by the end.  This
behaviour is clearly inconsistent with the hypothesis that the non-axisymmetric
structure in these more realistic models can be attributed to swing-amplified
particle noise. 

\eject

\ltwo{2.4 Coherent waves}
As the amplitude of the spiral waves in these experiments always reached levels
sufficient to heat the disc, the responsiveness and general level of
activity declined as the experiments evolved.  It was therefore impossible to
study structure in these uncooled models over long periods.  Therefore
Carlberg and I, in further (unpublished) work with this mass model, devised a
cooling algorithm which enabled us to run a long experiment that remained in a
quasi-steady state. 

Our cooling strategy in this case was to remove particles at random from the
disc and to re-insert them at some other randomly chosen position on a
locally-determined circular orbit.  The distribution from which the new
position was chosen was precisely that of the initial axisymmetric disc.  We
devised this cooling technique, which differs slightly from the accretion
method used by SC, in order to avoid a steadily rising disc mass.  This process
might be thought of as mimicking the death of a star and the formation of a new
star elsewhere in the disc -- the gas phase of the disc simply being omitted
from the dynamics.  It is physically unrealistic in an additional important
respect since it steadily redistributes angular momentum inwards, undoing that
transferred outwards by the spirals, in order to maintain a quasi-steady
distribution of angular momentum amongst the particles.  The cooling rate we
adopted was also quite unrealistically high -- 15 particles per time-step, or
15\% of the mass of the disc per rotation period. 

\topinsert{\hbox to \hsize{\hfill \figtwo \hfill}}\endinsert

We ran a $N=20$K model calculated according to this rule for 432 dynamical
times (one rotation period at the half-mass radius is 16 dynamical times). 
Starting from an initial $Q=1.7$, the model adjusted within the first $\sim 30$
dynamical times to quasi-steady values $1.8 \ltsim Q \ltsim 2.0$ over most
of the disc.  The $m=3$ component of $|A|$, from equation (1), is contoured in
Figure~2(a); the inclined stripes indicate recurrent shearing transient spiral
waves, rather similar to those reported in SC.  In this simulation, however,
the amplitude of successive events soon settles to an approximately constant
value, consistent with the quasi-steady state. 

Figure~2(b) shows the power spectrum of this apparently stochastic sequence of
spiral events.  Narrow horizontal ridges in this diagram indicate coherent
waves rotating at constant angular frequency over a wide range of inclination
angles.  The narrowness of the ridges in the frequency direction indicates that
the individual waves must be very long lived -- the ridges are scarcely broader
than our frequency resolution.

The spatial shapes of six of these apparently coherent waves are shown in
Figure~3.  All are trailing spiral patterns which show remarkable respect for
their principal resonances; the full-drawn circle marks the co-rotation
resonance, the dotted circles the inner and outer Lindblad resonances (ILR and
OLR respectively).  All but the most rapidly rotating of these long-lived
coherent waves possess ILRs, and only the slowest lacks an OLR within the
particle distribution.  The horizontal dashed lines in Figure~2(b) mark the
limits of the angular frequency range for which waves have both Lindblad
resonances within the disc of particles.

\pageinsert{\figthree}\endinsert

Though the waves survive for some time, they do not last indefinitely. 
Figure~2 shows data from the first half of the experiment -- a similar analysis
of the second half (not shown) reveals other long-lived waves, but their
frequencies, while in the same range as those shown in Figure~2(b), are not
identical.  Each wave appears to have a finite lifetime, which unfortunately is
not easy to measure -- a reasonable estimate might be $\sim 160$ dynamical
times or ten rotations.  New waves continually appear to replenish those that
decay. 

Long-lived coherent waves are hardly to be expected from swing-amplified
particle noise.  It is tempting to describe them as modes of oscillation of
the disc, though the fact that the majority possess ILRs is a real surprise.
Orthodox spiral mode theory (\eg\ Toomre 1981) might be able to account for
(a) with a feed-back loop and (f) as an edge mode (the horizontal dotted line
in Figure~3(b) shows the circular angular frequency at the outer edge) but it
could not account for most of the frequencies we observe, because such waves
should be strongly damped at both Lindblad resonances. 

\lone{3 Groove modes}
The instability described in this section arises when an otherwise smooth disc
has a deficiency of particles over a narrow range of angular momenta.  Further
details are given in Sellwood \& Kahn (1989).  If the disc is cold, \ie\ the
radial velocity dispersion $\sigma_u = 0$, then such a disc has a narrow,
axisymmetric groove in the surface density. 

\ltwo{3.1 Cold disc experiments}
We use the half-mass Mestel disc, which is the mass distribution of the Zang
disc described in \S2.1, but we start all the particles on exactly circular
orbits.  We suppress axisymmetric instabilities by softening the inter-particle
forces, preferring this approach to adding random motion only because the cause
the instability can be directly observed (Figure~6) and the analysis (\S3.2)
is more easily related to the experiments.  Discs with random motion should,
and do, behave similarly.  Moreover, the analysis remains valid for warm discs
as long as the disturbance has a wavelength much longer than the mean epicycle
size and one considers the distribution of guiding centres, instead of
particles. 

The surface density in our first experiment has no groove, and takes the form $$
\Sigma_0(r) = {qV_0^2 \over 2\pi Gr} T(r), \qquad {\rm with} \quad
T(r)=\cases{ [ 1 + (r/r_0)^4 ]^{-1}  & if $r<r_{\rm max}$; \cr
               0                    & otherwise. \cr}
$$ 
We set $q = 0.5$ (\ie\ a half-mass disc) and $r_{\rm max} = 9r_0$.  The
minimum softening length required to suppress axisymmetric instabilities in a
cold disc is $\epsilon_{\rm min} = \lambda_{\rm crit}/2\pi e$, where $e$ is the
base of natural logarithms.  We adopt a softening length of $1.5\epsilon_{\rm
min} \simeq 0.138r$, \ie\ increasing linearly with radius. 

We augment the central attraction of the model by an amount sufficient to
ensure centrifugal balance when all particles orbit at speed $V_0$.  This
central force, which can be thought of as arising from a halo, also corrects
the forces from the particle distribution for the central taper, outer cut-off
and softening.  The particles are smoothly distributed at the outset (\ie\ a
quiet start) and we restrict the disturbance forces to those arising from the
$m=2$ component of the surface density only, enabling us to confine the
particles to a semi-circle.  Again this restriction, which makes no difference
to the behaviour in the linear regime, is not a requirement of the computation.
 It does, however, further simplify the observed behaviour and considerably
reduces the number of particles we need employ. 

\topinsert{\hbox to \hsize{\hfill \figfour \hfill}}\endinsert

We adopt a system of units such that $r_0 = V_0 = G = 1$.  The orbital period
at $r=1$ is therefore $2\pi$ dynamical times. 

A simulation using just $15$K particles was run for 200 dynamical times during
which period no visible change occurred.  Fourier analysis revealed a very
slowly growing outer edge instability, but which was still well below its
saturation amplitude by the end of the test run.  We conclude therefore that
the basic cold disc model is stable enough to global $m=2$ perturbations for
our purposes. 

\pageinsert{\figfive}\endinsert

The next experiment we describe had almost the same initial particle
distribution, but differed by having a narrow, axisymmetric groove in the
surface density of the Lorentzian form $$ 
\Sigma(r) = \Sigma_0(r) \left[ 1 - {\beta w^2 \over (r-r_*)^2 + w^2} \right],
\eqno(2) 
$$ as illustrated in Figure~4.  We set the groove centre $r_* = 5$, its width
$w= 0.1$ and fractional depth $\beta = 0.8$.  Results from this simulation are
shown in Figure~5 -- the model is drastically unstable and develops a global,
large-amplitude, spiral pattern. 

The early evolution of the model is omitted from Figure~5 because no visible
changes occur, but Fourier analysis can detect a rapidly growing instability
from a very early stage.  The Fourier coefficients are extremely well fitted by
a single exponentially growing mode having the spatial form shown in Figure~6
and eigenfrequency $\omega = 0.383 + 0.072i$. The growth rate, $\Im(\omega)$,
is therefore more than one third the pattern speed, $\Re(\omega)/m$, indicative
of an extremely vigorous dynamical instability. 

\topinsert{\hbox to \hsize{\hfill \figsix \hfill}}\endinsert

Figure~6 shows that co-rotation (the full drawn circle) is close to, but just
outside the groove centre (marked by a dashed circle), and that the disturbance
is very strongly localised in this region.  There are weaker, trailing spiral
arms extending towards the Lindblad resonances (dotted circles) on either side
of the groove. 

\medskip
\ltwo{3.2 Local analytic treatment}
The global instability is driven by the non-axisymmetric density changes that
develop in a narrow range of radii around the groove.  We therefore find that a
local analysis of that region yields reasonable estimates of the mode
frequency. 

We use co-ordinates $(\xi,\eta)$ to represent the forced radial and transverse
displacements of a guiding centre from its circular orbit caused by the
presence of wave-like disturbance potential.  The origin of these displacement
co-ordinates moves on a circular path at radius $r$ with orbit speed $V_0$ and
therefore angular velocity $\Omega(r) \equiv V_0/r$. 

We consider a weak, uniformly rotating perturbing gravitational field with a
potential of the form $\Phi(r) e^{i(m\theta-\omega t)}$, which is small enough
for a linearised treatment of the forced motions to be valid. 

The linearised equations of motion are $$ \eqalign{
\ddot\xi + 2r\Omega\Omega^\prime\xi - 2\Omega\dot\eta &= {\partial\Phi \over
\partial r} \cr
\ddot\eta + 2\Omega\dot\xi &= {im\Phi \over r}, \cr} $$
with the exponential factor understood.  We have used a dot to denote
differentiation with respect to time (following the path of the guiding centre)
and a dash denotes differentiation with respect to $r$.  These equations can be 
readily integrated; setting $\kappa^2 = 2\Omega^2$ (for a flat rotation curve)
we obtain $$ 
\xi \sim - {m \over V_0 \kappa\nu} \Phi \qquad
{\rm and} \qquad \eta \sim {im \over r\kappa^2\nu^2} \Phi, \eqno(3)
$$ where $\nu = (\omega - m\Omega)/\kappa$, a dimensionless forcing frequency
that has a small real part near co-rotation.  We have therefore neglected terms
containing $\partial \Phi/\partial r$, since these enter with lower power of
$\nu$ in the denominator and are considerably smaller than the terms we retain.

The equation of continuity relates the disturbance surface density, $\Sigma_1$,
to the displacements of the guiding centres; \viz $$ 
\Sigma_1 = - {1\over r} {\partial \over \partial r} (\Sigma r \xi) -
{im \over r} \Sigma\eta. $$
In our current local treatment, in which $r$ is assumed large, we can neglect
curvature and use the approximate relation $$ 
\Sigma_1 \simeq - {\partial \over \partial r} (\Sigma \xi) - {im \over r}
\Sigma\eta. 
$$ We substitute from equations (3), and neglect radial derivatives of $\kappa$
and $\Phi$, which vary slowly across the narrow region around co-rotation; \ie\
the dominant radial derivatives are those of $\Sigma$ (which has a narrow
feature in the radial range of interest) and $1/\nu$ (which varies rapidly
because $\nu$ is small).  In this approximation, two of the terms cancel and we
are left with $$ 
\Sigma_1 \sim \Sigma^\prime {m \over V_0 \kappa \nu}\Phi. \eqno(4) $$ 

Equation (4) clearly gives a surface density of the form observed for the mode
(Figure~6).  The disturbed density is large where $\Sigma^\prime$ is large, but
the opposite signs of this gradient on each side of the groove cause a phase
shift of almost $\pi$ (\ie\ a 90\degrees\ shift in real space) across the
groove. 

Since we have neglected curvature, symmetry demands that co-rotation should
be at the centre of the groove, which determines the real part of the
eigenfrequency.  We now estimate the growth rate, $\omega_i$.

Because our softening length is large compared with the width of the groove, we
may compute the potential of the entire disturbance as if it were a single
sinusoidal wave on the circle at $r=r_*$ having an amplitude $$ 
C = \int \Sigma_1 dr 
  \simeq \int \Sigma^\prime {m \over V_0 \kappa \nu}\Phi dr, \eqno(5) 
$$ where the range of integration includes all stars whose orbits are
substantially perturbed.

We can write the surface density (2) in the vicinity of the groove in the form$$
\Sigma(x) = \Sigma_0(r) \left[ 1 - {\beta w^2 \over (r-r_*)^2 + w^2} \right]
\equiv {\Sigma_0(r) \over 2} \left[ 2 + i\beta w \left( {1 \over x - iw} -
{1 \over x + iw} \right)\right],
$$ where $x=r-r_*$.  Using this in (5), we find $$
C = {m \over V_0}\Phi \int_{-\infty}^\infty {-i\beta w\Sigma_0 \over 2} \left[
{1 \over (x-iw)^2} - {1 \over (x+iw)^2} \right] {1 \over \omega - m\Omega} dx. 
$$ Writing $\omega - m\Omega = mV_0(x-\lambda)/r_*^2$, where $\lambda=r_*
(1-\omega r_*/mV_0)$, we may evaluate the integral by closing off the contour
in the lower half of the complex plane (since $\Im(\lambda)<0$), and obtain $$
C = {m\Phi \over V_0} \cdot {-i\beta w\Sigma_0 r_*^2 \over 2mV_0} \cdot {-2\pi
i \over (\lambda - iw)^2}.
$$ Because co-rotation is at $r_*$, $\lambda = -i\omega_i r_*^2/mV_0$ (\ie\
purely imaginary).  The above expression therefore simplifies to $$
C = {\pi m^2\beta w\Sigma_0 r_*^2\Phi \over (\omega_ir_*^2 + mV_0w)^2}.
\eqno(6)
$$ 

The potential in the groove arising from this sinusoidally varying disturbance
with mass per unit length $C$ is
$$ 
\Phi = 2GCK_0 \left( {m\epsilon \over r_*} \right),
$$
where $K_0$ is the Bessel function of imaginary argument and $\epsilon$ is
the softening length.   Substituting this into (6), and setting $\Sigma_0 =
qV_0^2/2\pi Gr$, we obtain the desired expression for the growth-rate of the
mode
$$ 
\omega_i = {m V_0 w \over r_*^2} \left\{ \left[ {q\beta r_* \over w} K_0\left(
{m\epsilon \over r_*} \right) \right]^{1/2} - 1 \right\}. \eqno(7)
$$

This expression indicates that we expect instability only for grooves in which
the fractional depth
$$
\beta > {1 \over q K_0} {w \over r_*}, \eqno(8)
$$
\ie\ not for infinitesimally shallow grooves.  The critical depth is very
small, however; \eg\ in our experiments, $q =0.5$, $w=0.02r_*$ and $K_0 \simeq
1.45$ (for $m=2$), so we should expect an instability whenever the fractional
depth exceeds merely 2.8\%!  Notice also that the critical depth required
increases with the groove width (though our analysis assumes this to be small);
it would seem therefore, that the instability requires a critical density {\it
gradient\/} on the sides of the groove. 

We may compare the predictions of this local theory with the empirical results
from our experiments.  The eigenfrequency expected is $\omega_r = 0.4$ (for
co-rotation at the groove centre) and $\omega_i = 0.035$ [from (7)].  That
observed in the simulation is $\omega = 0.383 + 0.072i$.  The real parts are in
reasonable agreement, but our prediction considerably underestimates the
observed growth rate.  Two approximations in our analysis are largely
responsible for this poor agreement, the principal being the additional
contribution to $\Phi$ which comes from the supporting response of the disc.
Our neglect of curvature appears to be of lesser importance; in other
experiments, in which we raised the azimuthal mode number to $m=3$ and 4, where
it is more justifiable to neglect curvature, predictions for the growth rate
from equation (7) are somewhat closer to the observed values (and co-rotation
approaches the groove centre more closely still). 

But most of the growth rate discrepancy is removed only in an improved
treatment which includes the supporting response from the background disc
(Sellwood \& Kahn 1989).  The spiral response of the disc on either side of the
groove can be thought of as the polarisation response, or wake, induced by the
growing non-axisymmetric disturbance in the groove.  Discussions of wakes
usually focus on the steady response to a large co-orbiting disturbing mass
(\eg\ JT); in this case, we have an exponentially growing disturbance that
induces an exponentially growing response. 

The instability produces a growing sinusoidal distortion to each side of the
groove, and we must expect it to saturate once the distortions exceed the
groove width, \ie\ when further perturbations to the orbits of particles no
longer produce corresponding increases in disturbance forces.  At this stage,
the mass distribution at the radius of the original groove has large
non-axisymmetric variations -- to a good approximation, it will consist of two
blobs (for the $m=2$ component).  These large amplitude blobs will dissolve
only slowly and each will continue to induce a spiral wake for some
considerable time after the mode has saturated, as can be seen in Figure~5. 

As the clearest possible example of how untested theory can be completely
misleading, I note that the analysis provided here implies that a ridge, rather
than a groove, would not provoke instabilities.  As a ridge would be
characterised by a negative value for $\beta$, equation (8) indicates that no
instability should be expected.  A simple experimental test of this prediction
showed that it was completely wrong!  A ridge provokes instabilities just as
fierce as those from a groove.  We were able to understand ridge modes only
when we extended the analytic treatment presented here to consider wave-like
perturbations travelling around the ridge.  This more sophisticated treatment
is presented in Sellwood \& Kahn (1989) 

\medskip
\lone{4 A recurrent instability cycle}
In this section I argue that an experiment from my recent collaboration with
D. N. C. Lin provides a connecting link between the two previous sections.  Our
project was begun in the hope that $N$-body simulations of low mass particle
discs around a central point mass might reveal non-axisymmetric instabilities
of relevance to accretion discs.  The experiments did reveal non-axisymmetric
instabilities, but which are, however, likely to be of much greater
significance to spiral structure in galaxies than to instabilities in accretion
discs! 

The simulation, which is discussed in detail elsewhere (Sellwood \& Lin 1989),
revealed a recurrent instability cycle of the following form:  An instability
somewhere in the disc causes a non-axisymmetric wave to grow to large
amplitude.  Particles at the Lindblad resonances of the wave are strongly
scattered and change their angular momenta.  As the resonances are narrow,
the distribution function is severely depleted over a narrow range of
angular momentum, which leads to a new groove-type instability with co-rotation
at the radius of the Lindblad resonance of the previous wave.

\pageinsert{\figseven}\endinsert

Figure~7 shows how the distribution function changed in the experiment between
the start and a moment 267 dynamical times later.  The particle distribution is
shown in action space: the abscissa of each is its angular momentum and the
ordinate its radial action.  (Particles on more eccentric orbits have larger
radial actions.)  The arrows point to the principal resonances (co-rotation,
inner and outer Lindblad resonances) for the five waves observed in the
experiment up to the moment illustrated.  The resonant scattering that has
occurred has produced substantial heating (the particles have much larger
radial actions) and the distribution function is also much less smooth than at
the start.  Each tongue of particles reaching to large radial action was
produced by scattering at a Lindblad resonance. 

But the most interesting aspect of this revealing diagram is that the most
recent wave has left a clear deficiency of particles having angular momenta
$\sim 1.35$.  This deficiency drives a new instability in the disc which begins
its linear growth at about the time illustrated.  The new instability is quite
clearly a groove-type mode -- co-rotation lies at precisely the angular
momentum of the deficiency in the distribution function.  Note that in this
case we have a ``groove'' in the distribution of guiding centres; because
particles have random motion, there is no noticeable groove in the surface
density of the disc. 

\medskip
\lone{5 Discussion}
In this paper I have collected the evidence indicating that the recurrent
transient spirals seen in many simulations results from spiral {\it modes\/}
that grow rapidly and decay only to be replaced by new instabilities.  This
picture is radically different from two other current views of spiral
structure.  Bertin \etal\ (1989, and references therein) argue that spiral
patterns result from mild instabilities which lead to quasi-steady waves once
non-linear effects in the gas are taken into account.  It is not at all clear
that the models they propose could avoid the far more vigorous recurrent cycle
of groove modes presented here.  Toomre, on the other hand, stresses the role
of local spiral ``streaks'' resulting from density fluctuations in the disc
(\eg\ this conference).  Though these undoubtedly occur, it is far from clear
that real galaxy discs are ``lumpy'' enough for strong spirals to be produced
in this way. 

The original evidence for something more than noise came from the experiments
of SC.  The discussion of \S2 bolsters that case considerably by contrasting
the Sc models with the almost perfect illustration of swing-amplified
particle noise manifested by the Zang discs -- spiral amplitudes in the more
realistic Sc disc hardly changed with $N$.  Moreover, the recurring transient
patterns in cooled Sc models resulted from the super-position of a few
long-lived coherent waves.  Most of these waves, which could be picked out by
Fourier analysis, had two Lindblad resonances within the disc and therefore
could not be accounted for by orthodox mode theory. 

In the two subsequent sections I demonstrate that a groove-like feature in the
angular momentum distribution provokes a fierce instability and show how modes
of this type can recur in an instability cycle.  Resonant scattering by a large
amplitude wave, created by the saturation of one mode, carves a groove that
precipitates a new linear instability.  As the instability cycle also heats the
disc, making it less susceptible to further instabilities, activity must fade
after comparatively few cycles unless the disc is cooled.  The presence of a
dynamically significant fraction of gas in the galaxy is therefore essential
if the galaxy is to continue to display spiral structure over a long period. 

Though the connection still needs to demonstrated in massive discs, this
recurrent groove mode cycle seems at last to offer an explanation for the very
intriguing spiral activity that had been observed in simulations for many
years.   Experiments are currently in hand to attempt to substantiate this
claim. 

This paper has also brought out the importance of relating results from
simulations to the theory, and {\it vice-versa}.  Only through the quantitative
comparison of the results from the Zang model experiments with the predictions
of swing-amplified particle noise, could the more vigorous structure in the Sc
models be clearly identified as something more than noise.  On the other hand,
theory, unfettered by cross checks with experimental results, can easily go
astray.  An excellent example is provided by the beautifully straightforward
linear analysis, which gives a reasonable description of groove modes, but
which incorrectly predicts that axisymmetric density ridges in a disc are not
destabilising! 

\blankline
\noindent{\tenpoint Much of the work reported in \S2 was done in collaboration
with Ray Carlberg and has yet to be published.  Franz Kahn was my collaborator
for work described in \S3.  The author acknowledges the support of an SERC
Advanced Fellowship. \par} 

\def\apj{{\it Astrophys. J.}}
\def\araa{{\it Annu.~Rev.~Astron.~Astrophys.}}
\def\mnras{{\it Mon.~Not.~R. Astron.~Soc.}}
\lone{References}
{\tenpoint \parindent=0pt \everypar{\hangindent 1cm}\ignorespaces
Bertin, G., Lin, C. C., Lowe, S. A. \& Thurstans, R. P., 1989. \apj, {\bf 338},
pp.~78 \& 104. 

Goldreich, P. \& Lynden-Bell, D., 1965. \mnras, {\bf 130}, 124.

Julian, W. H. \& Toomre, A., 1966. \apj, {\bf 146}, 810 (JT).

Mestel, L., 1963. \mnras, {\bf 126}, 553.

Sellwood, J. A., 1986. In {\it The Use of Supercomputers in Stellar Dynamics},
Lecture Notes in Physics {\bf 267}, p.~5, eds. Hut, P. \& McMillan, S.,
Springer-Verlag, New York.

Sellwood, J. A., 1989. In {\it Nonlinear Phenomena in Vlasov Plasmas}, p. 87,
ed. Doveil, F., \'Editions de Physique, Orsay.

Sellwood, J. A. \& Carlberg, R. G., 1984. \apj, {\bf 282}, 61 (SC).

Sellwood, J. A. \& Lin, D. N. C., 1989. \mnras, in press.

Sellwood, J. A. \& Kahn, F. D., 1989. \mnras, in preparation.

Toomre, A., 1977. \araa, {\bf 15}, 437.

Toomre, A., 1981. In {\it Structure and Evolution of Normal Galaxies}, p.~111,
eds. Fall, S. M. \& Lynden-Bell, D., Cambridge University Press: Cambridge.

Zang, T. A., 1976. {\it PhD thesis}, MIT.

\par}
\end